# ON THE ASYMPTOTIC HYPERSTABILITY OF SWITCHED SYSTEMS

M. De la Sen*, A. Ibeas** and S. Alonso-Quesada*

*Institute of Research and Development of Processes, Faculty of Science and Technology
University of the Basque Country. PO. Box 644- Bilbao, Spain
Also with Department of Electricity and Electronics, Faculty of Science and Technology, University of the Basque Country. PO. Box 644- Bilbao, Spain. *Email(s):* manuel.delasen@ehu.es; santi@we.lc.ehu.es

** Department of Telecommunications and Systems Engineering
Universitat Autònoma de Barcelona, 08193- Bellaterra, Barcelona, Spain; *Email:* Asier.Ibeas@uab.cat

**Abstract**. Asymptotic hyperstability is achievable under certain switching laws if at least one of the feed-forward parameterization: 1) possesses a strictly positive real transfer function, 2) a minimum residence time interval is respected for each activation time interval of such a parameterization, and 3) a maximum allowable residence time interval is guaranteed for all active parameterization which is not positive real.

## 1. Introduction

The problems of hyperstability and asymptotic hyperstability have received important attention in Control Theory because global closed-loop stability is achieved for a wide class of nonlinear devices under the only constraint that they satisfy Popov´s-type integral inequalities including Lure´s and Popov´s absolute stability frameworks, [1-9]. It is needed that the linear feed-forward part of the system has a positive real transfer function for hyperstability and a strictly positive one for asymptotic hyperstability. In that way, global stability is achieved for a family of nonlinear controllers making the problem to be more independent of controller ageing or certain ranges of component dispersion along the fabrication process of the controller components. On the other hand, global stability of switched systems with several parameterizations has been investigated through exhaustive research work performed along the last years. In some cases, global asymptotic stabilization is achievable irrespective of the switching law, that is, for any sequence of switching time instants. This property is typically guaranteed for linear switched systems when all the parameterizations possess a common Lyapunov function, [10-14, 20]. Also, some results on stable parameterization switching in closed-loop time-varying systems based in theoretical considerations on operators in Hilbert spaces have been recently obtained in [35-41] which can be used by using techniques of non-periodic or adaptive sampling by respecting a minimum dwelling activation time among distinct parameterizations. The various stabilization techniques based on the concept of minimal residence time for switching in feedback linear systems have been studied a wide class of background literature as it has commented above in this section. In the general case, global stabilization of the switched system is achievable, depending on the switching time instants, provided that sufficiently large minimum time intervals are respected at certain stable active parameterizations. The minimum residence time intervals depend on the active system parameterizations from the last testing for a respected sufficiently large minimum residence time lower-bound, [15-22]. A similar problem has been investigated for a variety of linear time-varying systems like, for instance, delay-free systems, time-delay systems or hybrid systems, [11-14], [17-18], or for impulsive controls, [31]. There a basic technique consisting in respecting a minimum average dwelling time among the activation of the various parameterizations which has been exploited in a part of the background literature on the subject. The switched stabilization study is extended in such a paper to forced systems and to some classes of nonlinear systems under suitable uniformity assumptions. On the other hand, the topics of absolute stability and hyperstability are of a very relevant interest still nowadays because of their theoretical importance and its wide range of applications including stabilization under parametrical dispersion of regulator components either in the absence or in the presence of delays, hybrid mixed continuous- time and digital systems or passivity issues. See, for instance, some recent related background in [24-30] and references there in. Also, the property of absolute stability has been also investigated for systems involving time-delays. See, for instance, [32-33] and references therein. This paper relies on the asymptotic hyperstability of switched linear systems under regulation controls generated from nonlinear devices satisfying a class of Popov inequalities.

## 2. The closed-loop system

Consider the n- dimensional single-input single-output switched nonlinear feedback dynamic system whose structure is:

$$\dot{x}(t) = A_{\sigma(t)}(t)x(t) + b_{\sigma(t)}(t)u(t) \ ; \ x(0) = x_0 \in \mathbf{R}^n \tag{1}$$

$$u(t) = -\varphi_{\sigma_0(t)}(y(t), t); \ y(t) = c_{\sigma(t)}^T x(t) + d_{\sigma(t)} u(t) = c_{\sigma(t)}^T x(t) - d_{\sigma(t)} \varphi_{\sigma_0(t)}(y(t), t) \tag{2}$$

for $t \in \mathbf{R}_{0+} := \mathbf{R}_+ \cup \{0\}$, where $x(t) \in \mathbf{R}^n$, $u(t) \in \mathbf{R}$, $y(t) \in \mathbf{R}$ are, respectively, the state n- vector and scalar input, which is a feedback regulation control, and output where:

1) $\sigma : \mathbf{R}_{0+} \to \bar{p} := \{1, 2, ....., p\}$ and $\sigma_0 : \mathbf{R}_{0+} \to \bar{p}_0 := \{1, 2, ....., p_0\} \subset \bar{p}$, $p_0 \leq p$, for some given finite, and in general distinct, numbers $p, p_0 \in \mathbf{N}$ of parameterizations of (1)-(2). The first one describes a switching law among such various constant parameterizations defined by the set of quadruples $\{(A_i, b_i, c_i, d_i) : i \in \bar{p}\}$ of (1)-(2) of elements whose orders are compatible with the dimensionalities of the corresponding signals.

2) The function $\sigma : \mathbf{R}_{0+} \to \bar{p}$ is defined as $\sigma(t) = j = j(t) = j(t_i); \ \forall t \in [t_i, t_{i+1})$ for some integer $j \in \bar{p}$ for each $t_i \in \{t_i\}$, and each integer $i \in \bar{N} \subseteq \mathbf{N}_0 := \mathbf{N} \cup \{0\}$, where $STI = STI(\sigma) = \{t_i\}$ is the sequence of switching time instants generated from some given switching law $SL = SL(\sigma)$, subject to $t_{i+1} - t_i \geq T_r \geq 0; \ \forall i \in \mathbf{N}$. Such a function assigns at certain time intervals, a particular parameterization of the feed-forward part of the system which is modified at the switching time instants.

3) The function $\sigma_0 : \mathbf{R}_{0+} \to \bar{p}_0$ is defined as $\sigma_0(t) = j_0 = j_0(t) = j_0(t_{i_0}); \ \forall t \in [t_{i_0}, t_{i_0+1})$ so that $STI_0 = STI_0(\sigma_0) = \{t_{i_0}\} \subseteq STI \equiv \{t_i\}$, with the switching constraint $t \in STI_0 \Rightarrow t \in STI$, is the sequence of switching time instants of the feedback nonlinear part of the dynamic system and such a sequence is not necessarily identical to the switching sequence STI of the feed-forward linear part. That is, any switching time instant of the feedback part is always a switching time instant of the feed-forward part but the converse is not necessarily true. Such a function assigns a particular parameterization of the feedback nonlinear device of the system for certain time intervals which is modified at the switching time instants. This implies the following features:

a) The total number of parameterizations of the nonlinear feedback device function is at most that of the linear feed-forward block while it can be potentially smaller in the sense that a particular controller parameterization may be active along a time period including two or more switching actions on the linear feed-forward loop.

b) Even if the linear and nonlinear devices have the same number of parameterizations, the feedback switching activation of a new controller parameterization does not necessarily occur for any switching time instant of the feed-forward loop.

c) Any switching instant of the feedback nonlinear device is also a switching time instant of the linear feed-forward although the converse is not necessarily true. Furthermore, both switched parameterizations can have distinct ordering allocations within each sequence in the case that $\sigma_0 \neq \sigma$.

4) The nonnegative real number $T_r = T_r(\sigma)$ is the minimum residence (or dwelling) time interval at each parameterization and either $p = 1$ (i.e. the trivial case of a single parameterization of (1)-(2) with no switching) or $j(t_i) \neq j(t_{i+1})$ for all $t_i, t_{i+1} \in STI$. If $T_r = 0$ then the switching law is unconditional in the sense that switching is fully arbitrary. In the linear case, unconditional switching is possible in a stable way if all the parameterizations possess a common Lyapunov function. Otherwise, a minimum residence time $T_r > 0$ is required at each parameterization so as to guarantee the stabilization of the linear time-varying switched system if all of them are stable or at least one should be stable, subject to a minimum residence time when such a stable parameterization is active, which depends of the whole sequences and respective active time intervals at the rest of parameterizations.

5) The set $\overline{N}$ is a denumerable (proper or improper) subset of $N$ that, if finite, describes switching processes with a finite number of switches among the $p$ distinct parameterizations.

6) The control input is generated as the, in general, nonlinear output- feedback function $u(t) = -\varphi_{\sigma_0(t)}(y(t), t)$ which is assumed to be piecewise continuous while satisfying the integral Popov´s -type inequality:

$$\int_0^t \varphi_{\sigma_0(\tau)}(y(\tau), \tau) y(\tau) d\tau \geq -\gamma > -\infty \; ; \; \text{some } \gamma \in \mathbf{R}_+, \; \forall t \in \mathbf{R}_{0+} \tag{3}$$

Note that $\varphi : \overline{p}_0 \times \mathbf{R}^2 \to \mathbf{R}$ defined by $\varphi(t, y(t)) = \varphi_{\sigma_0(t)}(y(t), t)$ for $\sigma_0(t) = j_0 = j_0(t) = j_0(t_{i_0})$; $\forall t \in [t_{i_0}, t_{i_0+1})$ for some integer $j_0 \in \overline{p}_0$ for each $t_{i_0} \in \{t_{i_0}\}$ is piecewise continuous within its definition domain for any switching law $\sigma_0 : \mathbf{R}_{0+} \to \overline{p}_0 \subset \overline{p}$ if $\varphi_{j_0} : \mathbf{R}^2 \to \mathbf{R}$; $j_0 \in \overline{p}_0$ are all piecewise continuous. It is not required in principle that the nonlinear devices be distinct for each distinct parameterization in the feed-forward loop. That is, some of the controller nonlinear functions can be identical for different linear parameterizations of the feed-forward loop. The closed-loop system is displayed in Figure 1 below:

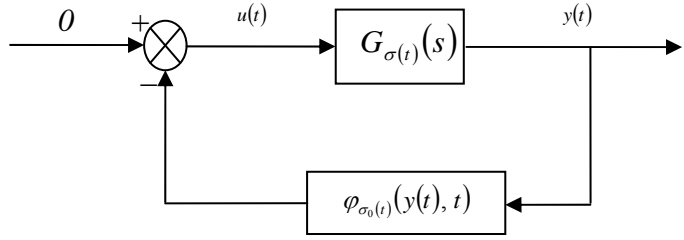

Figure 1. Block diagram of the feedback nonlinear system

Note that the switched feedback law $u(t) = -\varphi_{\sigma_0(t)}(y(t), t)$ together with (3) implies that (3) itself is equivalent to:

$$E(t) := \int_0^t y(\tau) u(\tau) d\tau \leq \gamma < \infty \ ; \ \text{some} \ \gamma \in \mathbf{R}_+, \ \forall t \in \mathbf{R}_{0+} \tag{4}$$

where $E(t)$ is an input-output energy measure of the feed-forward linear part of the closed-loop system.

## 3. Switching conditions for asymptotic convergence to zero of the input to the feed-forward loop

This section investigates parameterization switching sufficiency-type conditions for the input to the linear feed-forward loop to converge asymptotically to zero as time tends to infinity. The switching laws can involve parameterizations which are not strictly positive real being subject either a) to a finite number of switching actions; or b) to appropriate alternating with strictly positive real ones subject to maximum allowable residence times, or finally; c) to saturation- vanishing conditions of the input to the feed-forward linear loop. Note that the time-varying piecewise constant parameterization $\left( A_{\sigma(t)}, b_{\sigma(t)}, c_{\sigma(t)}, d_{\sigma(t)} \right)$ of the switched system (1)-(2) changes of values at time instants in $STI$. It is well-known that, in the absence of switching, i.e. if $p = 1$ then the closed-loop system is said to be hyperstable if the linear transfer function $G(s) = c^T (sI - A)^{-1} b + d$ is positive real, i.e. it belongs to the set $PR$ of positive real functions fulfilling $\operatorname{Re} G(s) \geq 0$ for $\operatorname{Re} s > 0$ (condition of hyperstability of the linear feed-forward subsystem) and the feedback law satisfies (3) (condition of hyperstability of the nonlinear feedback device). If $G(s)$ is strictly positive real, i.e. if it is in the set $SPR$ of strictly positive real functions fulfilling $\operatorname{Re} G(s) > 0$ for $\operatorname{Re} s \geq 0$ then the closed-loop system is said to be asymptotically hyperstable. It is well-known that realizable positive real functions are either stable or critically stable of relative order (or relative degree) either zero or one. Their critically stable poles, if any, are single and with non-negative associate residues. Strictly positive real transfer functions are, in particular, stable.

In order to simplify the formalism, we will refer indistinctly to positive realness and strict positive realness either for transfer functions or for their state-space realizations being in particular associated with the various parameterizations of the switched system. The feedback system is said to be hyperstable (respectively, asymptotically hyperstable) if it is globally stable for any nonlinear output- feedback law satisfying Popov´s inequality (3). For any given switching rule $\sigma : \mathbf{R}_{0+} \to \bar{p}$, let us consider the impulse response of the feed-forward linear block $g(t) := \mathbf{L}^{-1}(G(s))$; i.e. the Laplace anti-transform of $G(s)$, denote the Fourier transform of a function $f(t)$ as $F(f)$, provided that it exists, and also define the subsequent auxiliary truncated input it the switching action never ends:

$$u_{\sigma_0 (t_{i_0})}(t) = \begin{cases} u(t), & t \in [t_{i_0}, t_{i_0+1}) \\ 0, & t \in (-\infty, t_{i_0}) \cup [t_{i_0+1}, \infty) \end{cases} ; \forall t_{i_0} \in STI_0, \text{ for } i_0 = 0,1,\ldots,q_0(t)-1, \forall t \in \mathbf{R}$$

(5.a)

and

$$u_{\sigma_0 (t_{i_0})}(t) = \begin{cases} u(t), & t \in [t_{i_0}, t_{i_0+1}) \\ 0, & t \in (-\infty, t_{i_0}) \cup [t_{i_0+1}, \infty) \end{cases} ; \forall t_{i_0} \in STI_0, \text{ for } i_0 = 0,1,\ldots,q_0(t)-1, \forall t (\leq t_{q_0}) \in \mathbf{R} \quad (5.b)$$

$$u_{\sigma_0 (t_{q_0})}(t) = \begin{cases} u(t), & t \in [t_{q_0}, \infty) \\ 0, & t \in (-\infty, t_{q_0}) \end{cases} ; t_{q_0} = t_{q_0(t)} \in STI_0, \forall t (\geq t_{q_0}) \in \mathbf{R} \quad (5.c)$$

otherwise, i.e. if $t_{q_0(t)} = t_{q_0} < \infty$ is the last switching time instant (i.e. if $STI_0$ has a finite cardinal) under the axiom $t_0 (=0) \in STI \cap STI_o$ for any $\sigma : \mathbf{R}_{0+} \to \bar{p}$, $\sigma_0 : \mathbf{R}_{0+} \to \bar{p}_0$, and

$$q = q(t) := max\left(z \in \mathbf{N} \cup \{0\} : t_{q(t)} (\in STI) \leq t\right) ; \quad q_0 = q_0(t) := max\left(z \in \mathbf{N} \cup \{0\} : t_{q_0(t)} (\in STI_0) \leq t\right)$$

so that $t_q$ and $t_{q_0}$ are, respectively, the last switching time instants of the feed-forward linear parameterization and of the nonlinear feedback device in the time interval $[0, t]$ under the switching rule $\sigma : \mathbf{R}_{0+} \to \bar{p}$. In the same way, given $STI = \{t_i\}$, we define the impulse response as

$$g_\sigma(t) = g_\sigma(t_i); \forall t \in [t_i, t_{i+1})$$

If the convolution and Parseval's theorems are jointly applied to (4) for zero initial conditions and extending the definition of the input on $\mathbf{R}$ with $u(t) = 0$ and $g_\sigma(t) = 0$ for $t < 0$, one gets:

$$E(t) = \sum_{i=1}^{q(t)} \int_{t_{i-1}}^{t_i} y(\tau) u(\tau) d\tau + \int_{q(t)}^{t} y(\tau) u(\tau) d\tau$$

$$= \sum_{i=1}^{q(t)} \left( \int_{-\infty}^{\infty} \int_{-\infty}^{t_i} g_{\sigma(t_{i-1})}(\tau') u_{\sigma_0 (t_{i-1})}(\tau - \tau') u_{\sigma_0 (t_{i-1})}(\tau) d\tau' d\tau \right)$$

$$+ \int_{-\infty}^{\infty} \int_{-\infty}^{t} g_{\sigma(t_{q(t)})}(\tau') u_{\sigma_0 (t_{q(t)})}(\tau - \tau') u_{\sigma_0 (t_q)}(\tau) d\tau' d\tau$$

$$= \frac{1}{2\pi} \left( \sum_{i=1}^{q(t)} \int_{-\infty}^{\infty} Re\, G_{\sigma(t_{i-1})}(j\omega) \left| U_{\sigma_0 (t_{i-1})}(j\omega) \right|^2 d\omega + \int_{-\infty}^{\infty} Re\, G_{\sigma(t_{q(t)})}(j\omega) \left| U_{\sigma_0 (t_{q(t)})}(j\omega) \right|^2 d\omega \right) \quad (6)$$

; $\forall t \in \mathbf{R}_{0+}$ since $Im(j\omega) = -Im(-j\omega)$; $\forall \omega \in \mathbf{R}$ where $j = \sqrt{-1}$ is the complex unit. Now, proceed to calculate a lower-bound of (6) by using again Parseval's theorem and the hodograph symmetry property $Re(j\omega) = Re(-j\omega)$; $\forall \omega \in \mathbf{R}$ to yield:

$$E(t) \geq \frac{1}{2\pi} \left( \sum_{i=1}^{q(t)} \left( \min_{\omega \in \mathbf{R}_{0+}} Re\, G_{\sigma(t_{i-1})}(j\omega) \int_{-\infty}^{\infty} \left| U_{\sigma_0 (t_{i-1})}(j\omega) \right|^2 d\omega \right) + \min_{\omega \in \mathbf{R}_{0+}} Re\, G_{\sigma(t_{q(t)})}(j\omega) \int_{-\infty}^{\infty} \left| U_{\sigma_0 (t_{q(t)})}(j\omega) \right|^2 d\omega \right)$$

$$= \sum_{i=1}^{q(t)} \left( \min_{\omega \in \mathbf{R}_{0+}} Re\, G_{\sigma(t_{i-1})}(j\omega) \int_{-\infty}^{\infty} u^2_{\sigma_0(t_{i-1})}(\tau) d\tau \right) + \left( \min_{\omega \in \mathbf{R}_{0+}} Re\, G_{\sigma(t_{q(t)})}(j\omega) \right) \int_{-\infty}^{\infty} u^2_{\sigma_0(t_{q(t)})}(\tau) d\tau \quad (7)$$

; $\forall t \in \mathbf{R}_{0+}$. Now, consider the sequence of switching time instants until time $t$ of the given switching law $\sigma: \mathbf{R}_{0+} \to \overline{p}$, $STI(t) = STI \cap [0, t]$; $\forall t \in \mathbf{R}_{0+}$ and decompose it as the disjoint union $STI(t) = STI_p(t) \cup STI_n(t) \cup STI_z(t)$, $\forall t \in \mathbf{R}_{0+}$ as follows:

$$STI_p(\sigma, t) = STI_p(t) := \left\{ t_i \in STI(t) : \min_{\omega \in \mathbf{R}_{0+}} Re\, G_{\sigma(t_i)}(j\omega) > 0 \right\}$$

$$STI_n(\sigma, t) = STI_n(t) := \left\{ t_i \in STI(t) : \min_{\omega \in \mathbf{R}_{0+}} Re\, G_{\sigma(t_i)}(j\omega) = -\max_{\omega \in \mathbf{R}_{0+}} Re\, G_{\sigma(t_i)}(j\omega) < 0 \right\} \quad (8)$$

$$STI_z(\sigma, t) = STI_z(t) := \left\{ t_i \in STI(t) : \min_{\omega \in \mathbf{R}_{0+}} Re\, G_{\sigma(t_i)}(j\omega) = 0 \right\}$$

$STI_p = STI_p(\sigma)$, $STI_n = STI_n(\sigma)$ and $STI_z = STI_z(\sigma)$ are defined in the same way by including all the respective switching instants of the switching law $\sigma: \mathbf{R}_{0+} \to \overline{p}$, that is, the right-hand-side sets of the definitions in (8) modified for $t_i \in STI$. Note that, by technical reasons towards a clear proof of Lemma 1 below, (8) are defined so that $t \in STI \Rightarrow t \notin STI(t)$ following the definition convention $STI(t) = STI \cap [0, t]$. The first subsequent auxiliary result is concerned with the energy measure being nonnegative for all time for the switching law $\sigma: \mathbf{R}_{0+} \to \overline{p}$. The second one is referred to the non-negativity and boundedness of such a measure for all time.

**Lemma 1** (*non-negativity of the input-output energy*).
Define the switching-dependent amount:

$$g_\sigma(t_i) := \sum_{t_j \in STI_p(t_i)} \min_{\omega \in \mathbf{R}_{0+}} Re\, G_{\sigma(t_j)}(j\omega) \left( \int_0^{T_{j-1}} u^2(t_{j-1}+\tau) d\tau \right) - \sum_{t_j \in STI_n(t_i)} T_{j-1} \max_{\omega \in \mathbf{R}_{0+}} \left| Re\, G_{\sigma(t_j)}(j\omega) \right| \max_{t_{j-1} \leq \tau < t_j} u^2(\tau)$$

$$; \forall t_i \in STI \quad (9)$$

where $t_{j+1} = t_j + T_j$; $\forall t_j \in STI$ so that the $\sigma(t_j)$ parameterization of the feed-forward part of the system is active during a time interval $T_j$ in-between two consecutive switching time instants.
Then, the following properties hold:

**(i)** The input-output energy measure $E(t)$ is nonnegative for all time independent of the input $u: \mathbf{R}_{0+} \to \mathbf{R}$ if the switching law $\sigma: \mathbf{R}_{0+} \to \overline{p}$ satisfies any of the two conditions below:

  **(i.1)** All the active parameterizations are positive real.

  **(i.2)** Any active parameterization in an interval $[t_i, t_{i+1})$ which is not positive real is preceded by a strictly positive real one on $[t_{i-1}, t_i)$ while subject to a maximum (being potentially finite or infinity) residence time interval satisfying the constraint:

$$T_i = t_{i+1} - t_i < \frac{g_\sigma(t_i)}{\max_{\omega \in \mathbf{R}_{0+}} \left| Re\, G_{\sigma(t_i)}(j\omega) \right| \max_{0 \leq \tau \leq T_i} u^2(t_i + \tau)} \quad (10)$$

**(ii)** A necessary condition to guarantee that

$$g_\sigma(t) := g_\sigma(t_i) + \mu(t_i) \min_{\omega \in \mathbf{R}_{0+}} Re\, G_{\sigma(t_i)}(j\omega)\left(\int_{t_i}^{t} u^2(\tau)d\tau\right) - (1-\mu(t_i))(t-t_i) \max_{\omega \in \mathbf{R}_{0+}} \left|Re\, G_{\sigma(t_i)}(j\omega)\right| \max_{t_i \leq \tau < t} u^2(\tau)$$

$$; \forall t \in [t_i, t_{i+1}) \qquad (11)$$

is nonnegative is that the first parameterization after an arbitrary finite time is positive real in order to guarantee the non-negativity for all time of the input-output energy measure where $\mu: \mathbf{R}_{0+} \to \{0,1\}$ is a binary indicator function of value $\mu(t) = \mu(t_i) = 1$; $\forall t \in [t_i, t_{i+1})$ if $t_i \in STI_p \cup STI_z$ and $\mu(t) = \mu(t_i) = 0$ if $t_i \in STI_n$. A necessary condition to guarantee that (11) is nonnegative for all time, irrespective of the input, if the number of switches is finite is that the last active parameterization be positive real.

**(iii)** Assume that the first active parameterization after an arbitrary finite time is strictly positive real and that all non positive real parameterization, if any, satisfies the constraint of maximum residence time interval (10). Then, the input-output energy measure is positive for all $t > 0$.

**Proof**: Consider $g_\sigma(t)$ defined in (11). It turns out that $g_\sigma(0) = 0$. Also, $g_\sigma(t) \geq 0$; $\forall t \in [0, t_1]$ if $\sigma(0) \in [0, t_1)$ with $t_1 \in STI$; i.e. if the first active parameterization at $0 \leq t_0 < t_1 \leq \infty$ is positive real since

$$g_\sigma(t) := g_\sigma(t_0) + \mu(t_0) \min_{\omega \in \mathbf{R}_{0+}} Re\, G_{\sigma(t_0)}(j\omega)\left(\int_{t_0}^{t} u^2(\tau)d\tau\right) = \min_{\omega \in \mathbf{R}_{0+}} Re\, G_{\sigma(t_0)}(j\omega)\left(\int_{t_0}^{t} u^2(\tau)d\tau\right) \geq 0; \forall t \in [t_0, t_1]$$

(12)

Now, proceed by complete induction by assuming that $g_\sigma(\tau) \geq 0$ for $\tau \in [0, t_i]$ and any given $t_i \in STI$. Thus, $g_\sigma(t_i) \geq 0 \Rightarrow g_\sigma(t_i + \tau) \geq 0$ for any $\tau \in [0, T_i]$ by construction if: either

(a) $t_i \in STI_p \cup STI_z$ so that $g_\sigma(t_i) \geq 0$ so that $g_\sigma(t) \geq 0$; $\forall t \in [0, t_{i+1})$, then $t_{i+1} \in STI$ may be any positive arbitrary time instant; or

(b) $t_i \in STI_p \cup STI_z$ so that $g_\sigma(t_i) = 0$ implies that $g_\sigma(t) \geq 0$; $\forall t \in [0, t_{i+1})$ and then $t_{i+1} \in STI$ may be any positive arbitrary time instant ( note that this part of Case b is included in Case a) ; or

(c) $T_i = t_{i+1} - t_i < \dfrac{g_\sigma(t_i)}{\max_{\omega \in \mathbf{R}_{0+}} \left|Re\, G_{\sigma(t_i)}(j\omega)\right| \max_{0 \leq \tau \leq T_i} u^2(t_i + \tau)}$ if $g_\sigma(t_i) > 0$ and $t_i \in STI_n$ which is subject

to a maximum allowable guaranteed upper-bound except for the case of identically zero input on the current switching interval which allows an arbitrary next switching time instant $t_{i+1} \in STI_p \cup STI_z$ since

$t_{i+1} \in STI_n$ since the function $f(t_i, \delta) = \dfrac{g_\sigma(t_i)}{\max_{0 \leq \tau \leq \delta} u^2(t_i + \tau)}$ is nonnegative in the cases (b-c) (positive for the case c) and non-increasing, and then uniformly bounded on any definition domain $[0, \overline{T}_i]$ of nonzero measure, then there are always solutions in the incremental time argument $\delta$ that satisfy the constraint $0 \leq \delta \leq f(t_i, \delta)$. Thus, Property (i) follows for (i.1) from case a and it follows for (i.2) from cases b-c.

Property (ii) is proven as follows. Note that $g_\sigma(t) \geq g_\sigma(t_0) = g_\sigma(0) > 0$; $\forall t \in [t_0 = 0, t_1)$ for any $t_1 \in STI$ since the first active parameterization after an arbitrary finite time is strictly positive real. From the cases (a)-(b) of the proof of Property (i), it follows that

$$g_\sigma(t) \geq g_\sigma(t_0) = g_\sigma(0) > 0;\ \forall t \in [t_0 = 0, t_1) \Rightarrow g_\sigma(t) \geq g_\sigma(t_1) \geq g_\sigma(t_0) = g_\sigma(0) > 0;\ \forall t \in [t_0 = 0, t^*)$$

with $t^* \in STI$ being the first switching time instant activating a positive real parameterization on $[t^*, t^{**})$ with $t^{**}$ being the time instant of the next activation of a positive real parameterization. From the case (c) of the proof of Properties (i) for (i.1) and (i.2), it follows that $g_\sigma(t^*) > 0 \Rightarrow g_\sigma(t) > 0$; $\forall t \in [t_0 = 0, t^{**})$. The subsequent time intervals of alternate activation of positive real and nonpositive real parameterizations are discussed in the same way leading to a complete induction proof of Property (iii). The two necessary conditions of Property (iii) follow directly by using simple contradiction arguments. □

**Lemma 2** (*uniform boundedness of the input-output energy measure*). Assume that the switching law $\sigma: R_{0+} \to \bar{p}$ satisfies Lemma 1 and, furthermore, the nonlinear feedback device satisfies:

$$\int_{t_i}^{t_i+\eta} \varphi_{\sigma_0(\tau)}(y(\tau), \tau) y(\tau) d\tau \geq -\gamma - \int_0^{t_i} \varphi_{\sigma_0(\tau)}(y(\tau), \tau) y(\tau) d\tau;\ \forall t_i, t_{i+1} \in STI,\ \eta \in [0, T_i) \quad (13)$$

where $T_i = t_{i+1} - t_i$, and

$$\int_{t_q}^{t_q+\eta} \varphi_{\sigma_0(\tau)}(y(\tau), \tau) y(\tau) d\tau \geq -\gamma - \int_0^{t_q} \varphi_{\sigma_0(\tau)}(y(\tau), \tau) y(\tau) d\tau;\ \eta \in [0, \infty) \quad (14)$$

if $STI = \{t_0, t_1, ..., t_q\}$, $q < \infty$. Then, $0 \leq E(t) \leq \gamma < \infty$; $\forall t \in R_{0+}$, i.e. the input-output energy measure is nonnegative and uniformly bounded for all time.

**Proof**: It follows directly from Lemma 1, the fact that (13) and (14) guarantee Popov´s inequality (3) and the equivalence of (3) and (4). □

The following result gives conditions that guarantee that the input to the feed-forward loop is bounded and converges asymptotically to zero as time tends to infinity.

**Lemma 3**. Assume a switching law $\sigma: R_{0+} \to \bar{p}$ of the feed-forward loop such that all $\varphi_j: R^2 \to R$; $j \in \bar{p}_0$ are piecewise continuous for any active parameterizations while satisfying (13)-(14) guaranteeing the integral Popov´s inequality (3). Then, the following properties hold:

(**i**) If all the active parameterizations are strictly positive real then the input is uniformly bounded for all time and, furthermore, $\exists \lim_{t \to \infty} u(t) = 0$.

(**ii**) Assume that the first active parameterization after an arbitrary finite time is strictly positive real and that all active non positive real parameterization, if any, is preceded by a strictly positive real one while satisfying the constraint of maximum allowable residence time interval (10) for any active non positive real active parameterization. Then, the input to the feed-forward loop is uniformly bounded for all time and, furthermore, $\exists \lim_{t\to\infty} u(t) = 0$.

**(iii)** Assume that the switching law activates infinitely many times strictly positive real parameterizations and that first active parameterization after an arbitrary finite time is strictly positive real. Assume also that the system feed-forward loop of any active parameterization on $[t_i, t_{i+1})$ which is not positive real has no pole at $s=0$ and satisfies the saturation-vanishing input constraint $|u(t)| \leq K e^{-\lambda t_i}$; $\forall t \in [t_i, t_{i+1})$ with $t_i \in STI_n$ for some real constants $\lambda > 0$ and $K > 0$ subject to $\infty > \lambda > max\left(\lambda_0, \max_{t_i \in STI_n} \frac{\ln T_i}{2 T_i}\right)$ with $T_i = t_{i+1} - t_i$ for some prefixed $\lambda_0 \in \mathbf{R}_+$. Then, the input to the feed-forward loop is uniformly bounded for all time and, furthermore, $\exists \lim_{t\to\infty} u(t) = 0$.

**Proof**: (i) It all the parameterizations are strictly positive real then, one gets from (4), (9) and (12) since $STI(t) = STI_p(t)$; $\forall t \in \mathbf{R}_{0+}$:

$$0 \leq g_\sigma(t_i) := \sum_{t_j \in STI(t_i)} \min_{\omega \in \mathbf{R}_{0+}} Re\, G_{\sigma(t_j)}(j\omega)\left(\int_0^{T_{j-1}} u^2(t_{j-1} + \tau)d\tau\right) \leq \gamma < \infty; \quad \forall t_i \in STI \qquad (15)$$

$$0 \leq \sum_{t_i \in STI} \min_{\omega \in \mathbf{R}_{0+}} Re\, G_{\sigma(t_i)}(j\omega)\left(\int_{t_i}^{t_{i+1}} u^2(\tau)d\tau\right) \leq \liminf_{t\to\infty} g_\sigma(t) \leq \limsup_{t\to\infty} g_\sigma(t) \leq \gamma < \infty$$

$$0 \leq g_\sigma(t) := g_\sigma(t_q) + \min_{\omega \in \mathbf{R}_{0+}} Re\, G_{\sigma(t_q)}(j\omega)\left(\int_{t_q}^{t} u^2(\tau)d\tau\right) \leq \gamma < \infty, \quad \forall t \geq t_q (\in STI) \text{ if } STI \cap (t_q, \infty) = \emptyset \qquad (16)$$

The last property implies that there is a finite number of active parameterizations, all being strictly positive real. Note that all of them are finite for all $s \in \mathbf{C}$ since strictly positive real transfer functions cannot possess critical poles, then they are integrators-free. First, assume that $\min_{\omega \in \mathbf{R}_{0+}} Re\, G_{\sigma(t)} \geq d > 0$; i.e. the relative order of the strictly positive real transfer function is zero so that it has the same number of zeros and poles. Note that the above amounts are, furthermore, strictly bounded from below by zero if the input $u: \mathbf{R}_{0+} \to \mathbf{R}$ is piecewise continuous and non-identically zero. Furthermore, if the switching action never ends then $card(STI) = \infty$ and $t_i (\in STI) \to \infty$. Thus, $\lim_{t_i (\in STI) \to \infty} \int_{t_i}^{t_i + \tau} u^2(\tau) d\tau = 0$; $\forall \tau \in [0, T_i)$ and since the input is piecewise continuous and the sequence of integrals $\left\{\int_{t_i}^{t_i + \tau} u^2(\tau) d\tau\right\}$ has infinitely many elements, it follows that $\exists \lim_{t\to\infty} u(t) = 0$ since $\min_{t_i \in STI, \omega \in \mathbf{R}_{0+}} Re\, G_{\sigma(t_i)} > 0$ (otherwise the sum of infinitely may elements in (15) can not be bounded which leads to a contradiction). If switching ends in a finite time instant $t_q \in STI$ for some $q \in \mathbf{N}_0$ then $card(STI) = q + 1 < \infty$ and there is no switching for $t > t_q$ so that $\lim_{t\to\infty} \int_{t_q}^{t} u^2(\tau) d\tau = 0$ and, again, $\exists \lim_{t\to\infty} u(t) = 0$ since $\int_{t_q}^{t} u^2(\tau) d\tau$ is a strictly increasing function of time for any $0 \leq t_q < \infty$ and any nonzero input, then contradicting (16), unless $\exists \lim_{t\to\infty} u(t) = 0$. Next, assume that $\min_{\omega(<\infty) \in \mathbf{R}_{0+}} Re\, G_{\sigma(t)} > 0$ and $\lim_{\omega \to \infty} Re\, G_{\sigma(t)} = 0$. In this case, define a strictly decreasing nonnegative real function $\omega_0 : \mathbf{R}_{0+} \to \mathbf{R}_{0+}$, that is, $\omega_0 = \omega_0(\varepsilon) \to +\infty$ as $\varepsilon \to 0^+$. Thus, (15) leads to:

$$0 \leq \min_{t_j \in STI, \omega \in \mathbf{R}_{0+}} Re\, G_{\sigma(t_j)}(j\omega) \sum_{t_j \in STI(t_i)} \left( \int_{t_{j-1}}^{t_j} u^2(\tau) d\tau \right)$$

$$\leq \gamma_0 - 2 \sum_{t_j \in STI(t_i)} \int_{\omega_0(\varepsilon)}^{\infty} Re\, G_{\sigma(t_j)}(j\omega) \left| U_{\sigma_0(t_{i-1})}(j\omega) \right|^2 d\omega < \infty \,;\, \forall t_i \in STI \qquad (17)$$

for any positive finite real constants $\varepsilon, \varsigma$, and $\gamma_0$ so that

$$\infty > \gamma_0 > \liminf_{\varepsilon \to 0^+} \left( \gamma + \varsigma + 2 \sum_{t_j \in STI(t_i)} \int_{\omega_0(\varepsilon)}^{\infty} Re\, G_{\sigma(t_j)}(j\omega) \left| U_{\sigma_0(t_{i-1})}(j\omega) \right|^2 d\omega \right) = \gamma + \varsigma > 0$$

since Popov´s inequality (3) implies that $\int_0^t \varphi_{\sigma_0(\tau)}(\tau) y(\tau) d\tau \geq -\gamma \geq -\gamma_0 > -\infty$ for any finite real constant $\gamma_0 \geq \gamma$ and $\lim_{\varepsilon \to 0} \omega_0(\varepsilon) = +\infty$. Thus, $\exists \lim_{t \to \infty} u(t) = 0$ remains valid if the strictly positive real transfer functions have relative order equal to one. Property (i) has been proven.

Property (ii) follows from similar considerations as those in the proof of Property (i) from (9) and (12) under the maximum allowable residence time constraint (10) for any active non positive real active parameterizations provided that the first active parameterization after an arbitrary finite time is strictly positive real.

Property (iii) is proven by noting that: 1) $t_i \in STI_z$ does not need to be accounted for in Popov´s inequality or equivalent feed-forward loop since the minimum real value of its associate feed-forward transfer function is zero; 2) one gets the following relations for parameterizations which are not positive real under no critical pole at $s = 0$, what implies $\left| Re\, G_{\sigma(t_i)}(j\omega) \right| < \infty$ together with the input saturation –vanishing constraint $|u(t)| \leq K e^{-\lambda t_i}$; $\forall t \in [t_i, t_{i+1})$ for $t_i \in STI_n$:

$$-\sum_{t_i \in STI_n} \int_{t_i}^{t_{i+1}} y(\tau) \varphi_{\sigma_0(t_i)}(y(\tau), \tau) d\tau = \sum_{t_i \in STI_n} \int_{t_i}^{t_{i+1}} u(\tau) y(\tau) d\tau \leq \frac{1}{2\pi} \sum_{t_i \in STI_n} \max_{\omega \in \mathbf{R}_{0+}} \left| Re\, G_{\sigma(t_i)}(j\omega) \right| \int_{-\infty}^{\infty} U_{\sigma_0(t_i)}(j\omega) d\omega$$

$$\leq K^2 \max_{\omega \in \mathbf{R}_{0+}, t_i \in STI_n} \left| Re\, G_{\sigma(t_i)}(j\omega) \right| \sum_{t_i \in STI_n} T_i e^{-2\lambda t_i} \leq \frac{K'^2}{1 - e^{-2\lambda_0 t^*}} < \infty \qquad (18)$$

Since a (finite) $\lambda$ - constant fulfilling $\infty > \lambda > \max\left( \lambda_0, \max_{t_i \in STI_n} \frac{\ln T_i}{2 T_i} \right)$ exists since $\max_{t_i \in STI_n} \frac{\ln T_i}{2 T_i} < \infty$, equivalently, $\max_{t_i \in STI_n} \left( T_i e^{-2\lambda T_i} \right) < 1$, for $T_i \in [0, \infty)$ and $\limsup_{T_i \to \infty} \max\left( \lambda_0, \max_{t_i \in STI_n} \frac{\ln T_i}{2 T_i} \right) < \infty$. where $t^* = \min(t : t \in STI_n)$. The combination of (15)-(17) with Popov´s inequality equivalent versions (3) and (4), guaranteed under (13)-(14), Lemma 2 and (6) yields after separating the bounded contribution of non strictly positive real parameterizations from the strictly positive real ones of zero relative order:

$$0 < \min_{t_i \in STI_p} G_{\sigma(t_i)} \int_0^{\infty} u^2(\tau) \mu(\tau) d\tau < E(t) \leq \gamma + \frac{K'^2}{1 - e^{-2\lambda_0 t^*}} < \infty \qquad (19)$$

Since the switching law possesses infinitely many strictly positive real active parameterizations by hypothesis, it follows that $\int_0^t u^2(\tau)\mu(\tau)d\tau$ cannot be a strictly increasing function of time, since, otherwise, a contradiction would follow from (19), thus, $\exists \lim_{t\to\infty} u(t)=0$. For the case of relative order equal to one, the proof follows closely to the corresponding case in the proof of Property (i). □

## 4. Asymptotic hyperstability

The following known result for linear switched systems will be used. It is known for switchings among stable parameterizations that if a set of matrices of dynamics has a common Lyapunov function then stability is preserved under arbitrary switching.

**Theorem 1**. The following properties hold:

**(i)** Let $A = \{A_i : i \in \overline{p}\}$ be a set of $p$ Hurwitz matrices. Then all matrices in $\mathbf{A}$ have a common Lyapunov function if and only if $\sum_{i=1}^{p}(A_i X_i + X_i A_i^T) < 0$ for any given n- matrices $X_i \geq 0$; $i \in \overline{p}$ ("$\geq$" denoting matrix positive semidefiniteness). The open-loop switched system (1), i.e. $u \equiv 0$, is globally asymptotically stable for any given arbitrary switching law $\sigma: \mathbf{R}_{0+} \to \overline{p}$.

**(ii)** The matrices in $\mathbf{A}$ do not have a common Lyapunov function if and only if there is a set of n-matrices $X_i \geq 0$; $i \in \overline{p}$ such that $\sum_{i=1}^{p}(A_i X_i + X_i A_i^T) = 0$.

**(iii)** The matrices in $\mathbf{A}$ do not have a common Lyapunov function if at least one non-Hurwitz matrix exists in the set $\overline{A} = A \cup \{A_i^{-1} : i \in \overline{p}\}$.

**(iv)** The set $A = \{A_i : i \in \overline{p}\}$ of Hurwitz matrices has a common Lyapunov function only if $\sum_{i=1}^{p}(\alpha_i A_i + \beta_i A_i^{-1})$ is a Hurwitz matrix; $\forall \alpha_i, \beta_i \in \mathbf{R}_{0+}$ such that $\sum_{i=1}^{p}(\alpha_i A_i + \beta_i A_i^{-1}) \in \mathbf{R}_+^{n \times n}$.

**(v)** The set $A = \{A_i : i \in \overline{p}\}$ of Hurwitz matrices has a common Lyapunov function $V(x(t)) = x^T(t)Px(t)$ where $P = P^T > 0$ is a positive definite real n-matrix if

$$\|A_i - A_k\|_2 < \frac{\lambda_{min}(Q_k)(|\lambda_{max}(A_k)| - \varepsilon)}{K_k^2 \lambda_{max}(Q_k)} = \frac{\lambda_{max}(A_k^T P + PA_k)(|\lambda_{max}(A_k)| - \varepsilon)}{K_k^2 \lambda_{min}(A_k^T P + PA_k)} \qquad (20)$$

; $\forall i \in \overline{p}$ for any given $k \in \overline{p}$, for any given arbitrary real constant $\varepsilon \in (0, |\lambda_{max}(A_k)|)$ and some testable real constant $K_k \geq 1$, where $\lambda_{max}(.)$ and $\lambda_{min}(.)$ stand for the (.)-matrix and $Q_k = -(A_k^T P + PA_k)$. The open-loop system (1)-(2) is globally exponentially stable for any arbitrary switching law $\sigma: \mathbf{R}_{0+} \to \overline{p}$, $\forall p \in \mathbf{N}$.

**Proof**: Properties (i)-(iv) are given in [23-24]. Property (v) is proven as follows: Since $A_k$ is Hurwitz then it satisfies a Lyapunov equation $A_k^T P + PA_k = -Q_k = -Q_k^T < 0$ for any given $k \in \overline{p}$ and $Q_k = Q_k^T > 0$ and $P = P^T = \int_0^\infty e^{A_k^T \tau} Q_k e^{A_k \tau} d\tau > 0$ satisfying the Lyapunov equation since for any real constant $\varepsilon \in \left(0, |\lambda_{max}(A_k)|\right)$ ($\varepsilon \in \left[0, |\lambda_{max}(A_k)|\right]$ if all the eigenvalues of $A_k$ are distinct) and some testable real constant $K_k \geq 1$, one has:

$$\|P\|_2 = \lambda_{max}(P) = \left\|\int_0^\infty e^{A_k^T \tau} Q_k e^{A_k \tau} d\tau\right\|_2 \leq \frac{K_k^2 \lambda_{max}(Q_k)}{2(|\lambda_{max}(A_k)| - \varepsilon)} \tag{21}$$

Then,

$$A_i^T P + PA_i = A_k^T P + PA_k + \left(A_i^T - A_k^T\right)P + P(A_i - A_k) = -Q_i = -Q_i^T < 0 \; ; \; \forall i \in \bar{p} \tag{22}$$

is guaranteed if

$$2\lambda_{max}(P)\|A_i - A_k\|_2 < \lambda_{min}(Q_k) \Leftrightarrow \|A_i - A_k\|_2 < \frac{\lambda_{min}(Q_k)(|\lambda_{max}(A_k)| - \varepsilon)}{K_k^2 \lambda_{max}(Q_k)} \; ; \; \forall i \in \bar{p} \tag{23}$$

and any given $j \in \bar{p}$ and Property (v) follows since $\dot{V}(x(t)) = -x^T(t)Q_{\sigma(t)}(t)x(t) < 0$ if $x(t) \neq 0$ with $Q_{\sigma(t)}(t) = Q_k > 0$ for some $k \in \bar{p}$; $\forall t \in \mathbf{R}_{0+}$ and any given switching law $\sigma: \mathbf{R}_{0+} \to \bar{p}$. □

The following result basically establishes in its simpler form that, if there is at least a stable parameterization and the residence time is subject to a minimum residence for each time interval where such a parameterization is active, then the open-loop system (1) is globally exponentially stable.

**Theorem 2**. Let $\{\lambda_{ij} : j \in \bar{n}\}$ be the spectrum of $A_i$; $\forall i \in \bar{p}$ and let $K_i(\in \mathbf{R}_+) \geq 1$, $\rho_i \in \mathbf{R}$; $i \in \bar{p}$ be constants such that:

$\max_{j \in \bar{n}} Re \, \lambda_{ij} \leq -\rho_i$ ($\max_{j \in \bar{n}} Re \, \lambda_{ij} < -\rho_i$ if there is some eigenvalue of $A_i$ of multiplicity larger than one for any $i \in \bar{p}$), and $\left\|e^{A_i t}\right\| \leq K_i e^{-\rho_i t}$; $\forall i \in \bar{p}$.

Consider the open-loop system with a switching law $\sigma: \mathbf{R}_{0+} \to \bar{p}$ generating the following sequence of switching time instants:

$$STI = STI(\sigma) = \left\{t_{i_0}, t_{i_0+1}, \dots, t_{i_0+i_1} = t_1^*, t_{i_0+i_1+1}, \dots, t_{i_0+i_1+i_2} = t_2^*, \dots, t_{\sum_{j=0}^k i_j} = t_k^*, \dots\right\} \tag{24}$$

with $\{i_j\}$ being a finite or infinite strictly increasing sequence of nonnegative integers subject to $i_0 = 0$ and $|i_{j+1} - i_j| \leq \xi < \infty$ under the following assumptions:

**A.1**. The set $A$ contains at least one Hurwitz matrix.

**A.2**. $STI \supset STI^* = \{t_i^* : i \in \bar{N}^* \cup \{0\}\}$ for some $\bar{N}^* = \{1, 2, \dots N^*\} \subset \mathbf{N}$ is a set of marked switching time instants chosen provided that either a minimum residence time constraint is guaranteed if $N^* = \infty$ for each current marked active parameterization $\sigma(t_i^*)$; $\forall i \in \bar{N}^* \cup \{0\}$, being necessarily stable, according to:

$$T^*_{\sigma\left(t_{\sum_{j=0}^k i_j}+\ell\right)} \geq \frac{1}{\rho}\left(\sum_{\ell=0}^{i_{k+1}-1}\left|\ln K_{\sigma\left(t_{\sum_{j=0}^k i_j}+\ell\right)}\right| - \sum_{\ell=1}^{i_{k+1}-1}\rho_{\sigma\left(t_{\sum_{j=0}^k i_j}+\ell\right)}T_{\sigma\left(t_{\sum_{j=0}^k i_j}+\ell\right)} - |\ln \delta|\right) \tag{25}$$

for some real constant $\delta \in (0,1)$, or $N^* < \infty$ with the last active parameterization $[t_{N^*}, \infty)$ being Hurwitz. Thus, the open-loop switched system is globally exponentially stable.

**Proof**. Note that the marked switching time instants $STI^* = \{t_i^* : i \in \overline{N}^* \cup \{0\}\}$ and their corresponding residence time instants are related with the values of $STI$ as follows:

$$t_{k+1}^* - t_k^* = t_{\sum_{j=0}^{k+1} i_j} - t_k^* = t_{\sum_{j=0}^{k+1} i_j} - t_{\sum_{j=0}^{k} i_j} = \sum_{\ell=0}^{i_{k+1}-1} T^*_{\sum_{j=0}^{k} i_j + \ell} \; ; \; T_k^* = t_{\sum_{j=0}^{k} i_j + 1} - t_k^* = t_{\sum_{j=0}^{k} i_j + 1} - t_{\sum_{j=0}^{k} i_j}$$

$$; \forall t_i \in STI, \; \forall t_k^* \left(= t_{\sum_{j=0}^{k} i_j}\right) \in STI^* \tag{26}$$

Note that for identically zero input the unforced solution of (1) satisfies on the time interval $\left[t_k^* = t_{\sum_{j=0}^{k} i_j}, t_{k+1}^* = t_{\sum_{j=0}^{k+1} i_j}\right)$ under some active stable parameterization with a Hurwitz matrix of dynamics $A_{\sigma\left(t_{\sum_{j=0}^{k} i_j}\right)} \in A = \{A_i : i \in \overline{p}\}$ leading to $x(t_k^* + \tau) = e^{A_{\sigma\left(t_{\sum_{j=0}^{k} i_j}\right)} \tau} x(t_k^*)$ ;

$\forall \tau \in \left[0, T^*_{\sum_{j=0}^{k} i_k}\right)$ and the stability abscissa of the Hurwitz matrix $A_{\sigma\left(t_{\sum_{j=0}^{k} i_j}\right)}$ being

$-\left|\rho_{\sigma\left(t_{\sum_{j=0}^{k} i_j}\right)}\right| < 0$. Thus, one gets the following condition if the marked active residence time within $\left[t_k^*, t_{k+1}^*\right)$ is large enough to satisfy the minimum residence time constraint (25):

$$\frac{\|x(t_{k+1}^*)\|}{\|x(t_k^*)\|} \leq \left\| e^{\sum_{\ell=1}^{i_{k+1}-1} \rho_{\sigma\left(t_{\sum_{j=0}^{k} i_j + \ell}\right)} T^*_{\sigma\left(t_{\sum_{j=0}^{k} i_j + \ell}\right)}} \left(\prod_{\ell=0}^{i_{k+1}-1} K_{\sigma\left(t_{\sum_{j=0}^{k} i_j + \ell}\right)}\right) e^{-\left|\rho_{\sigma\left(t_{\sum_{j=0}^{k} i_j}\right)}\right| T^*_{\sigma\left(t_{\sum_{j=0}^{k} i_j + \ell}\right)}} \right\| \leq \delta < 1 \tag{26}$$

and for any bounded initial conditions, the sequence $\{\|x(t_k^*)\|\}$ converges exponentially to zero for identically zero control provided that there are infinitely many switches ( i.e. if switching does not end in finite time so that $\overline{N}^* = \infty$). Also, $x(t_k^* + \tau) = e^{A_{\sigma\left(t_{\sum_{j=0}^{k} i_j}\right)} \tau} x(t_k^*) \to 0$ as $x(t_k^*) \to 0$ ;

$\forall \tau \in \left[0, T^*_{\sum_{j=0}^{k} i_k}\right)$. Note also that $x(t_{N^*}^* + \tau) = e^{A_{\sigma\left(t_{\sum_{j=0}^{k} i_j}\right)} \tau} x(t_{N^*}^*) \to 0$ at exponential rate as $\tau \to \infty$ is bounded for any bounded initial conditions since $t_{N^*}^* < \infty$, since $N^* < \infty$, implies that $x(t_{N^*}^*)$ is bounded. Thus, the open-loop switched system is globally exponentially stable. □

**Theorem 3**. Consider a switched feedback system (1)-(2) with the nonlinear feedback device satisfying a Popov´s inequality constraint (13), eventually both constraints (13)-(14) in the case of a finite number of switches. The following properties hold:

**(i)** Assume that all the transfer functions of the feed-forward loop being built with the various matrices of the set $A$ are strictly positive real (then also Hurwitz) and, furthermore, (20) holds, under the necessary and sufficient condition $\sum_{i=1}^{p}(A_i X_i + X_i A_i^T) < 0$ for any given n-matrices $X_i \geq 0$; $i \in \bar{p}$. Then, the closed-loop system is asymptotically hyperstable for any switching law $\sigma: \mathbf{R}_{0+} \to \bar{p}$, for any $p \in \mathbf{N}$, such that the nonlinear feedback device satisfies Lemma 2, i.e. Popov´s inequality (13), eventually (13)-(14) if it generates only a finite number of switches. If, in particular, $\varphi_{\sigma(t)}(t) = \varphi(t)$; $\forall t \in \mathbf{R}_{0+}$, i.e. the nonlinear device does not depend on switching and satisfies Popov´s inequality (3), then the closed-loop systems is unconditionally asymptotically hyperstable for any arbitrary switching law $\sigma: \mathbf{R}_{0+} \to \bar{p}$.

**(ii)** Assume that all the matrices of $A$ are Hurwitz with their associate transfer functions being strictly positive real. If $\sum_{i=1}^{p}(A_i X_i + X_i A_i^T) = 0$ for some n-matrices $X_i \geq 0$, $i \in \bar{p}$, or if $\sum_{i=1}^{p}(\alpha_i A_i + \beta_i A_i^{-1})$ is not Hurwitz for some $\alpha_i$, $\beta_i \in \mathbf{R}_{0+}$ subject to $\sum_{i=1}^{p}(\alpha_i A_i + \beta_i A_i^{-1}) \in \mathbf{R}_+^{n \times n}$, then the closed-loop switched system (1)-(2) is not unconditionally asymptotically hyperstable for a switching-independent feedback nonlinear device $\varphi_{\sigma(t)}(t) = \varphi(t)$; $\forall t \in \mathbf{R}_{0+}$ any given switching law $\sigma: \mathbf{R}_{0+} \to \bar{p}$ even if the transfer functions of all the parameterizations of the feed-forward loop are strictly positive real.

**(iii)** Consider a switched system (1)-(2) possessing (at least) one strictly positive real parameterization of the feed-forward loop under some switching law $\sigma: \mathbf{R}_{0+} \to \bar{p}$ whose associate set of switching time instants $STI = STI(\sigma)$ is subject to the subsequent constraints:

1) The first active parameterization through time after an arbitrary finite time is strictly positive real.

2) All active non positive real parameterization, if any, satisfies the constraint of *maximum allowable residence time constraint* (10), via (9), and it is preceded by a strictly positive real one.

3) There is a set $STI^* \subset STI$ of marked switching time instants for some of the active strictly positive real parameterizations satisfying the *minimum residence time constraint* (25) for the time interval $[t_k^*, t_{k+1}^*)$ defined for each two consecutive marked switching instants.

4) The nonlinear feedback device satisfies Popov´s inequality (13) for the set $STI$, eventually (13)-(14) if it generates only a finite number of switches.

Then, the closed-loop system is asymptotically hyperstable for such a switching law $\sigma: \mathbf{R}_{0+} \to \bar{p}$. □